\documentclass[10pt,conference]{IEEEtran}
\usepackage{wrapfig}

\usepackage{algorithm2e} 
\usepackage{amsmath}
\usepackage[american]{babel}
\usepackage{booktabs}
\usepackage{caption}
\usepackage{color}
\usepackage{comment}
\usepackage[keeplastbox]{flushend}
\usepackage[acronym]{glossaries}
\usepackage{listings}
\usepackage{multirow}
\usepackage{rotating}
\usepackage{siunitx}
\usepackage{soul}
\usepackage{tabularx}
\usepackage{tikz}
\usepackage[textsize=scriptsize]{todonotes}
\usepackage{url}
\usepackage{xspace}
\usepackage{graphicx}
\usepackage[export]{adjustbox}
\usepackage{subfigure}
\usepackage{enumitem}
\usepackage{authblk}

\usepackage{hyperref}

\newcommand{\toolname}{\textit{hybrid-Falcon}\xspace}
\newcommand{\tooldataset}{\textit{AndroNetMnist}\xspace}

\begin{document}

\title{\toolname : Hybrid Pattern Malware Detection and Categorization with Network Traffic and Program Code}

\author[1]{Peng Xu}
\author[1]{Claudia Eckert}
\author[2]{Apostolis Zarras}

\affil[1]{Technical University of Munich}
\affil[2]{Delft University of Technology}

\maketitle

\begin{abstract}
Nowadays, Android is the most dominant operating system in the mobile ecosystem, with billions of people using its apps daily. As expected, this trend did not go unnoticed by miscreants, and Android became the favorite platform for discovering new victims through malicious apps. Moreover, these apps have become so sophisticated that they can bypass anti-malware measures to protect the users. Therefore, it is safe to admit that traditional anti-malware techniques have become cumbersome, sparking the urge to develop an efficient way to detect Android malware. 
	
This paper presents ~\toolname, a hybrid pattern Android malware detection and categorization framework. It combines dynamic and static features of Android malware, which are from network traffic and code graph structure. In~\toolname, we treat network traffic as a dynamic feature and process it as a 2D-image sequence. Meanwhile, ~\toolname handles each network flow in the packet as a 2D image and uses a bidirectional LSTM network to process those 2D-image sequences to obtain vectors representing network packets. We use the program code graph for a static feature and introduce natural language processing (NLP) inspired techniques on function call graph (FCG). We design a graph neural network-based approach to convert the whole graph structure of Android apps to vectors. Finally, We utilize those converted vectors, both network and program code features, and concatenate them to detect and categorize the malware. Our results reveal that ~\toolname yields better results as we get 97.16\% accuracy on average for malware detection and 88.32\% accuracy for malware categorization. Additionally, we release a dataset~\tooldataset, which converts the network traffic to a 2D-image sequence and helps to accomplish malware detection on a 2D-image sequence. 

\end{abstract}

\section{Introduction}
Android increasingly becomes the most popular mobile operating system in the world. Unfortunately, due to its reputation, it has also become the leading target platform for attackers. Adversaries use Android to launch millions of malicious apps that dupe victims into revealing their private data or performing malicious operations, such as spying on users' actions, propagating spam, or launching unwanted advertisements. At the same time, Android malware investigation, which includes malware detection and categorization, has become a crucial task for security investigators in both industry and academia. As a result, numerous research works have attempted to detect Android malware~\cite{canfora2015effectiveness,gascon2013structural,li2018android,mclaughlin2017deep,hybroid}. 

Recently, a significant portion of the proposed approaches leverages the contextual information of Android applications to detect and categorize Android Malware, such as the permission and sensitive application programming interfaces (APIs). For example, Li et al.~\cite{li2018android} presented a classifier using the factorization machine (FM) architecture, where they extract various Android app features (permission, malicious APIs, etc.) from manifest files and source code. Similarly, Chen et al.~\cite{chen2018android} proposed an approach that analyzes Android malware based on its static behavior that involves the use of permissions, components, and sensitive APIs calls. Meanwhile, there are also several works using natural language processing(NLP)~\cite{kang2016n,xu2021detecting} based methods to detect Android Malware, and those systems consider Android apps similar to documents and instructions in Android apps as words. Take ~\cite{kang2016n} as an example, and it evaluates the effectiveness of uni-gram, bi-gram, and tri-gram with stacked generalization.

Additionally, several existing works are based on the Android network traffic pattern to detect the legitimate and malicious behaviors~\cite{malik2016credroid,arora2014malware,zulkifli2018android,marin2020deepmal,wang2017malware,xu2021hawkeye,xu2020manis,xu2021falcon}. Most of them are based on the manual indicated rules and build rule-based classifiers for detecting Android malware, such as CICAndMal2017~\cite{lashkari2018toward}, which uses the third-party tool to extract nine features from the network PCAP (packet capture) file and detect malware with those exact features. Zulkifli et al. ~\cite{zulkifli2018android} presents a decision tree algorithm on network traffic as dynamic features of Android apps.  

Although the above methods serve the Android platform with an extra security layer, they have limitations. On the one hand, for instance, the contextual information of program code struggles against malware obfuscation procedures, such as ~\textit{identifier renaming}, ~\textit{string encryption}, ~\textit{dead-code insertion}, and ~\textit{instruction substitution}. These procedures cause changes in the compiled code to evade detection easily. For example, using string encryption, the FM method in Li et al.~\cite{li2018android} would be influenced because the system cannot extract the available permission information from the manifest file. Meanwhile, other NLP-based malware detection systems~\cite{kang2016n,xu2021detecting} are also influenced by some obfuscation techniques.

%
On the other hand, for the network traffic-based methods, the biggest problem is the code coverage problem~\cite{tazaki2013direct}. As one type of dynamic feature-based method, the network traffic only represents a few of code functionality. Most of the code functionalities do not need or generate network traffic. On the other hand, most of the state-of-the-art network traffic-based malware detection systems are rule-based methods~\cite{malik2016credroid,arora2014malware,zulkifli2018android,marin2020deepmal,wang2017malware,hybroid} , which sophisticated attacks can easily evade since rule-based analysis relies on distinguishing expected versus anomalous behavior; these methods may suffer when malware is modified to hide its footprints or behavior. 

Therefore, in this paper, we present a hybrid pattern malware detection and categorization framework that combines the network and program code features. First, we convert the network packets to the sequence of 2D gray images and leverage a convolution neural network (CNN) to pre-train the 2D gray images. We then use a bi-directional LSTM network to process the continuous network traffic and do the malware classification similar to the 2D image sequence classification task. 

Additionally, we design an NLP inspired-method for the program code pattern based on the graph neural network (GNN). It can detect obfuscated applications while maintaining excellent performance. In brief, we first design the \texttt{opcode2vec}, \texttt{function2vec}, and \texttt{graph2vec} components to represent instruction, function, and the whole program's information with vectors. After obtaining features both from the network traffic and program code, we then feed these vectors into the classifier and train it to differentiate between benign and malicious applications.  Finally, \toolname identifies the Android malware families and shows that it outperforms most existing frameworks. We get 97.16\% accuracy for malware detection and 88.32\% accuracy for malware categorization average.
In this paper, we highlight the novelty from two parts. The first one is the bi-directional LSTM to process the 2D-image sequence. And the other is graph-based malware detection, which takes natural language processing techniques as a node representation method.

\bigskip \noindent In summary, we make the following main contributions:

\begin{itemize}
	\item We introduce \toolname, a hybrid pattern Android malware detection and categorization framework that combines the network traffic and program code features.
	\item We design a bidirectional LSTM network to accomplish 2D gray images sequence classification, which takes the network packets (converted to 2D images) as input.
	\item We create and release a dataset, \tooldataset, which includes 3,255,391 2D gray images in five categories(Benign, Adware, Ransomware, Scareware and SMSmalware) for network traffic classification.
	\item We design and implement an NLP-inspired malware detection and categorization module for the program code that can process obfuscated programs.
	\item We evaluate the accuracy of our approach using real malicious and benign datasets.
	\item We release our work as an opensource project after the paper will be accepted. 
\end{itemize}

\section{Background}
\subsection{Android Applications}
Android applications are written in Java and executed within Dalvik virtual machine, a customized Java VM, or ART (Android Runtime). Android applications consist of many components of various types, which are the basic modules of the application. The basic elements of any Android application include activities, intent and broadcast receivers, services, content providers, widgets, and notifications. The application's permission is to protect the privacy of Android users. Android applications must request permission to access sensitive user data such as Contacts and SMS, as well as certain system features such as Camera and Internet. Depending on the functionality, the system might grant permission automatically or might prompt the user to approve the request.

A DEX file that includes the bytecode of Android Dalvik instruction can be transformed into SMALI files, where each file represents a single class and contains its methods. Each method includes instructions, and each instruction consists of a single opcode and multiple operands. In total, Dalvik provides 256 opcodes, but some of them are reserved for future usage. Right now, we use 221 opcodes in real Android applications.

\subsection{Android malware detection}
With the increasing popularity of Android smartphones in recent years, the topic of detecting Android malware and categorizing its families attracts several researchers' attention. As every malware detection system, the Android malware detection can be classified into two types: the traditional feature codes based method, and the machine learning based method. Regarding the conventional feature codes based approach, the detector checks the classic malicious behaviors, such as memory corruption, buffer overflow as well as return-oriented programming (ROP)~\cite{sun2016blender} code reuse attacks. In machine learning based method, we can get the features through static~\cite{li2018android,gascon2013structural,mclaughlin2017deep,canfora2015effectiveness}, and dynamic code analysis~\cite{yan2012droidscope,tam2015copperdroid,wong2016intellidroid}. Additionally, some works use a hybrid approach, as they combine static and dynamic analysis together~\cite{zhao2014attack,xu2016hadm}. 

Additionally, to reduce the application's reverse-engineering, several obfuscation techniques have increasingly been introduced into Android applications to render the source code into random and unreadable files to either protect the ownership of the code or easily evade detection in case of malware.

\subsection{Natural Language Processing}
\label{w2v}
In natural language processing, to perform text sentiment, document classification and machine translation using machine learning or neural network, there exists several approaches to transform the text information to vectors, namely word~\cite{mikolov2013distributed},  sentence~\cite{lin2017structured,palangi2016deep,conneau2017supervised,pagliardini2017unsupervised} and document~\cite{le2014distributed} embeddings. 

Word2vec~\cite{mikolov2013distributed} is an embedding method that transforms words into embedding vectors. Similar words should have similar embeddings. Word2vec uses the skip-gram network, which is the neural network with one hidden layer. The skip-gram is trained to predict the neighbor word in the sentence.

Sentence2vec~\cite{lin2017structured,palangi2016deep,conneau2017supervised,pagliardini2017unsupervised} is an extension of word2vec to sentences. The sentence embedding is considered to be the average of the source word embeddings of its elemental words. Sentence2vec has supervised version~\cite{lin2017structured,palangi2016deep,conneau2017supervised} and also unsupervised version~\cite{pagliardini2017unsupervised}. 

Doc2vec~\cite{le2014distributed} modifies the word2vec algorithm to the unsupervised learning of continuous representations for larger blocks of text, such as sentences, paragraphs, or entire documents. PV-DM model of doc2vec acts as a memory that remembers what is missing from the current context. While the word vectors represent the concept of a word, the document vector intends to represent the concept of a document.  
\subsection{Graph embedding}
Graphs are meaningful and understandable representations of data. Graph embedding~\cite{goyal2018graph,perozzi2014deepwalk,grover2016node2vec,kipf2016semi,wang2016structural} is the transformation of property graphs to a vector or a set of vectors. Graph embedding should capture the graph topology, vertex-to-vertex relationship, and other relevant information about graphs, subgraph, and vertices. Further, the classification systems on graph structures, which are based on the graph embedding, are defined as the graph-based classification.

\section{Motivation}

\section{System Design and Implementation}
\subsection{Overview}
\begin{figure*}
	\centering
	\includegraphics[width=1\textwidth]{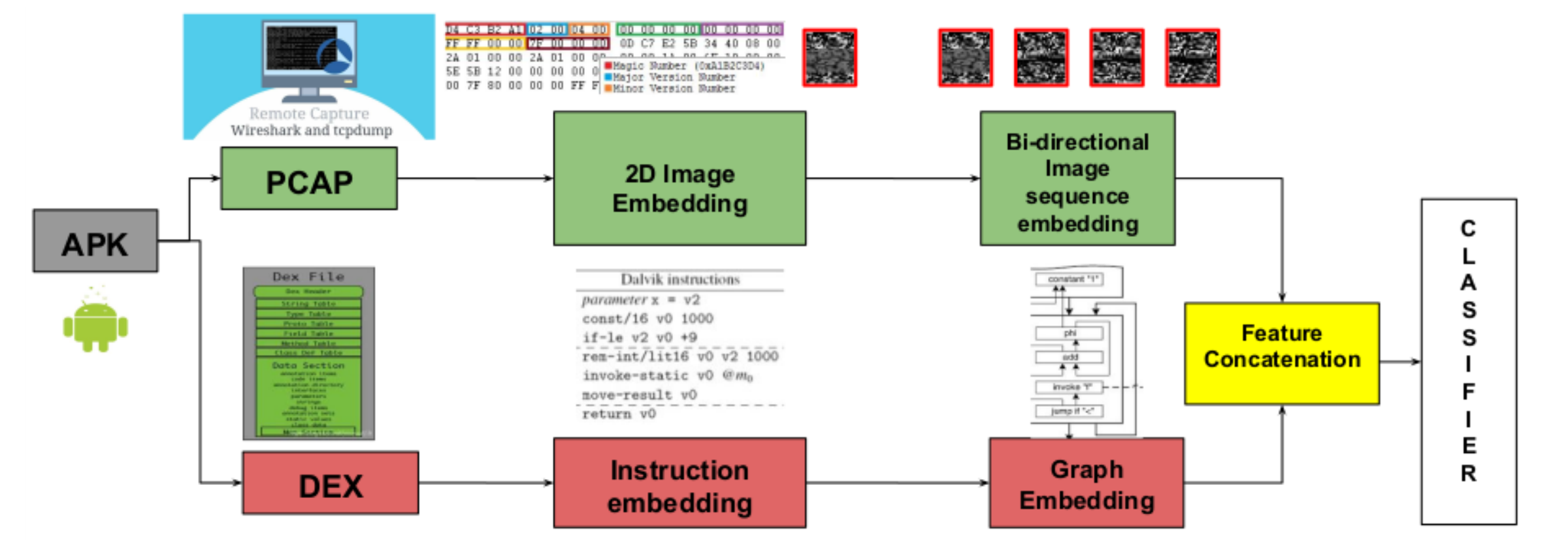}
	\caption{The architecture of \toolname }
	\label{fig:architect}
\end{figure*}

The architecture of \toolname is presented in Figure~\ref{fig:architect}. Our hybrid pattern malware detection and categorization framework includes two group of features and their corresponding modules. 

For the network traffic based module, we input the \textit{PCAP (packet capture)}~\footnote{https://wiki.wireshark.org/SampleCaptures} file and convert each network flow in PCAP file to a 2D image and pre-train model on 2D images with CNN network. We use the pre-trained model to convert each 2D image to a vector and process the continuous network flows in PCAP file as a 2D-image sequence by a bi-directional LSTM network. We will present this part in Section~\ref{network} in detail. The reason why we consider to convert network flow to 2D image and the whole network packet as 2D-image sequence is that extract network feature automatically by representation learning method, rather than manual indicated methods, such as port-feature, source and target IP address and other limited features which are indicated by human. 

For the program code (DEX file in Android APK), we extract function call graphs (FCGs) from DEX~\footnote{https://source.android.com/devices/tech/dalvik/dex-format} file. We use the pre-trained natural language process inspired instruction/opcode embedding and function embedding methods to convert each node of FCGs (the nodes are functions in the program) to vectors. We use graph embedding to convert each graph to vector. We will present this part in Section~\ref{code} in detail. 

Finally, we combine converted vectors both from network traffic and program code into a combined feature and feed it into the classifier (various classifiers are evaluated). 
\subsection{Features from Network Traffic}
\label{network}This section presents our method to convert network traffic to vectors based on image classification and transfer learning. The architecture is presented in Figure~\ref{fig:network}. To compare to other works in this field, we have two challenges. The first challenge \textbf{C1} is how to classify each network flows (several network packets), and the second challenge \textbf{C2} is how to classify the whole network packets based on the split flows. 
\subsubsection{Network Packet and Flow}
\begin{figure}
	\centering
	\includegraphics[width=0.45\textwidth]{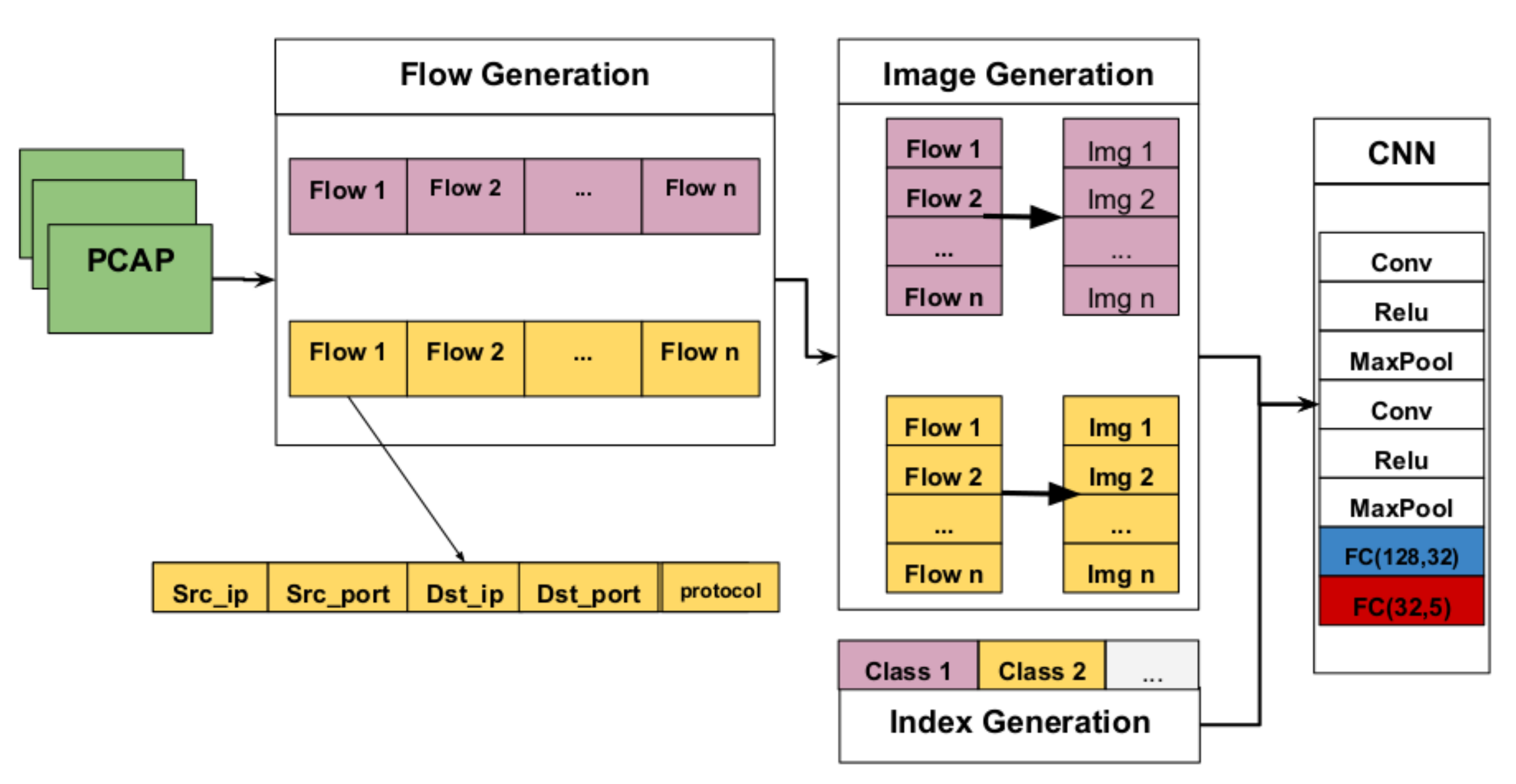}
	\caption{Converting Network Traffic to Vectors}
	\label{fig:network}
\end{figure}
For the network traffic analysis, there are three different granularity, raw packet level, flow level, and session level~\cite{wang2017malware}. 
In our work, we take the network flow as our analysis target. All raw packets from PCAP files are defined as a set $P$ =$\{p^1,p^2,\cdots,p^{|P|}\}$, and every packet is defined as $p^i$ = ($x^{i},b^{i},t^{i}$), where $i = 1,2, \cdots, |P|$ and $x^{i}$ stands for a 5-tuple, which includes source IP, source port, destination IP, destination port and the protocol types (e.g., TCP, UDP), where $b^{i}$ and $t^{i}$ stand for the packet's size and the starting time of the packet, respectively. Network flow groups several packets that have the same 5-tuple. In this way, the challenge \textbf{C1} is solved.  Meanwhile, for the network flow level analysis, it is shown as the flow generation in Figure~\ref{fig:network}. All raw packets in the same network flow are arranged in time order. 
\subsubsection{Network Flow to Image}
\label{img_gen}
As we mentioned in the above section, we split the network flow from the raw network packets. After getting network flow files, we convert them to 2D images like the image generation in Figure~\ref{fig:network}. Here we utilize the trimming and padding methods to normalize all network flows that have the same size. In~\toolname we set the 784 bytes as our size, and it includes the 5-tuple and the following bytes in the network flow. If network flow' size is larger than 784 bytes, we trim it to 784 bytes. If the size of those flow files' is smaller than 784 bytes, we pad them by 0x00 to 784 bytes. Finally, we convert those trimmed and padded files to 2D gray images. Each byte of the original file represents a pixel, such as 0x80 is gray, and 0xff is white.  

We also generate the class label in this step, which stands for the different network traffic classes. In our work, we define five-different labels because we have four various malware families and one benign group. That is reasonable for our malware categorization task. We pre-train the model indirectly for our malware detection task, which is based on the previous malware categorization. After categorizing all samples into five groups (four malware groups and one benign group. Then we add all malware in four different groups as the total number of malware samples.
\subsubsection{Transfer Learning and Feature Generation}
\label{img2vec}
In our work we leverage a 8-layer convolution neural network to pre-train our converted 2D gray images. Our model has 70,213 total parameters. After the previous step, we transform our malware categorization and detection tasks into a 5-category classification problem.

\begin{equation}
\begin{split}
Y^1 = MaxPooling_{2*2}(Relu(conv2d_{3*3}(X_{28 * 28}))) \\
Y^2 = MaxPooling_{2*2}(Relu(conv2d_{3*3}(Y^1))) \\
Y^3 = FC_{128,32}(Y^2) \\
Y = FC_{32,5}(Y^3) 
\end{split}
\end{equation}
We use our 5-categories(Benign, Adware, Ransomware, Scareware and SMSmalware) classification task to train the model. After getting the pre-train model, we take $Y^3$ that has 32-bit vector as our features for the next step.
We use \textit{sparse\_categorical\_crossentropy} loss and Adam optimizer and set the learning\_rate as 0.001 and epoch as 50. We use one dropout layer between MaxPool2 and FC1, and we set the dropout rate as 0.5.

\begin{algorithm}
	\scriptsize
	\DontPrintSemicolon 
	\KwIn{Raw network traffic files(PCAPs), Classes of raw network traffic}
	\KwOut{Network traffic embedding ${v_{t} : t \in N}$}
	\textit{I. Pre-train 2D classification model, extract image sequence with batch\_size = 8 and map 2D images to vectors}
	
	\For{n=1 to N}{
		\For{$v \in n$}{
			\textcolor{blue}{//Forward LSTM}
			
			$i_{t_{f}} = \alpha(x_{t_{f}}*U_{f}^{i} + h_{t-1_{f}}*W_{f}^{i})$
			
			$f_{t_{f}} = \alpha(x_{t_{f}}*U_{f}^{f} + h_{t-1_{f}}*W_{f}^{f})$
			
			$o_{t_{f}} = \alpha(x_{t_{f}}*U_{f}^{f} + h_{t-1_{o}}*W_{f}^{o})$
			
			$\tilde{C_{t_{f}}} = tanh(x_{t_{f}}*U_{f}^{g} + h_{t-1_{f}}*W_{f}^{g})$
			
			$C_{t_{f}} = \alpha (f_{t_{f}} * C_{t-1_{f}} + i_{t_{f}} * \tilde{C_{t_{f}}})$
			
			$h_{t_{f}} = tanh(C_{t_{f}}) * o_{t_{f}}$
			
			\textcolor{blue}{//Backward LSTM}
			
			$i_{t_{b}} = \alpha(x_{t_{b}}*U_{b}^{i} + h_{t-1_{b}}*W_{b}^{i})$
			
			$f_{t_{f}} = \alpha(x_{t_{b}}*U_{b}^{f} + h_{t-1_{b}}*W_{b}^{f})$
			
			$o_{t_{b}} = \alpha(x_{t_{b}}*U_{b}^{f} + h_{t-1_{o}}*W_{b}^{o})$
			
			$\tilde{C_{t_{b}}} = tanh(x_{t_{b}}*U_{b}^{g} + h_{t-1_{b}}*W_{b}^{g})$
			
			$C_{t_{b}} = \alpha (f_{t_{b}} * C_{t-1_{b}} + i_{t_{b}} * \tilde{C_{t_{b}}})$
			
			$h_{t_{b}} = tanh(C_{t_{b}}) * o_{t_{b}}$
		}
		\textbf{$h_{t} = concat(h_{t_{f}}, h_{t_{b}})$}
	}
	
	\Return{$v_f$}
	\caption{Continuous Network Flow Embedding}
	\label{algo:lstm}
\end{algorithm}
\subsubsection{Continuous Network Traffic Processing}
So far, we convert the network flows to images and pre-train the 2D gray image-based multi-class classification model. However, we cannot classify those malware or benign samples based on each 2D gray image for our malware detection and categorization task. Normally, each PCAP file includes hundreds or thousands of raw network packets and network flows. Therefore, the malware classification issue converts to a continuous 2D image classification task. In other words, that is a 2D-image sequence classification or sequential image classification~\cite{li2018independently,bai2018trellis}. Most of the sequential image classification works combine the RNN and CNN and put RNN focusing on the sequential task and CNN for the image features. Meanwhile, in the natural language processing (NLP) field, in order to process the sequential issues with the pre-trained model, BERT~\cite{devlin2018bert}, GPT (v2,v3)~\cite{radford2018improving,budzianowski2019hello} and other transformers(e.g., ELMo~\cite{peters2018deep}, Transformer~\cite{vaswani2017attention}) are introduced into to capture the sequence relationship by leveraging the LSTM or RNN networks. 

\begin{figure}
	\centering
	\includegraphics[width=0.51\textwidth]{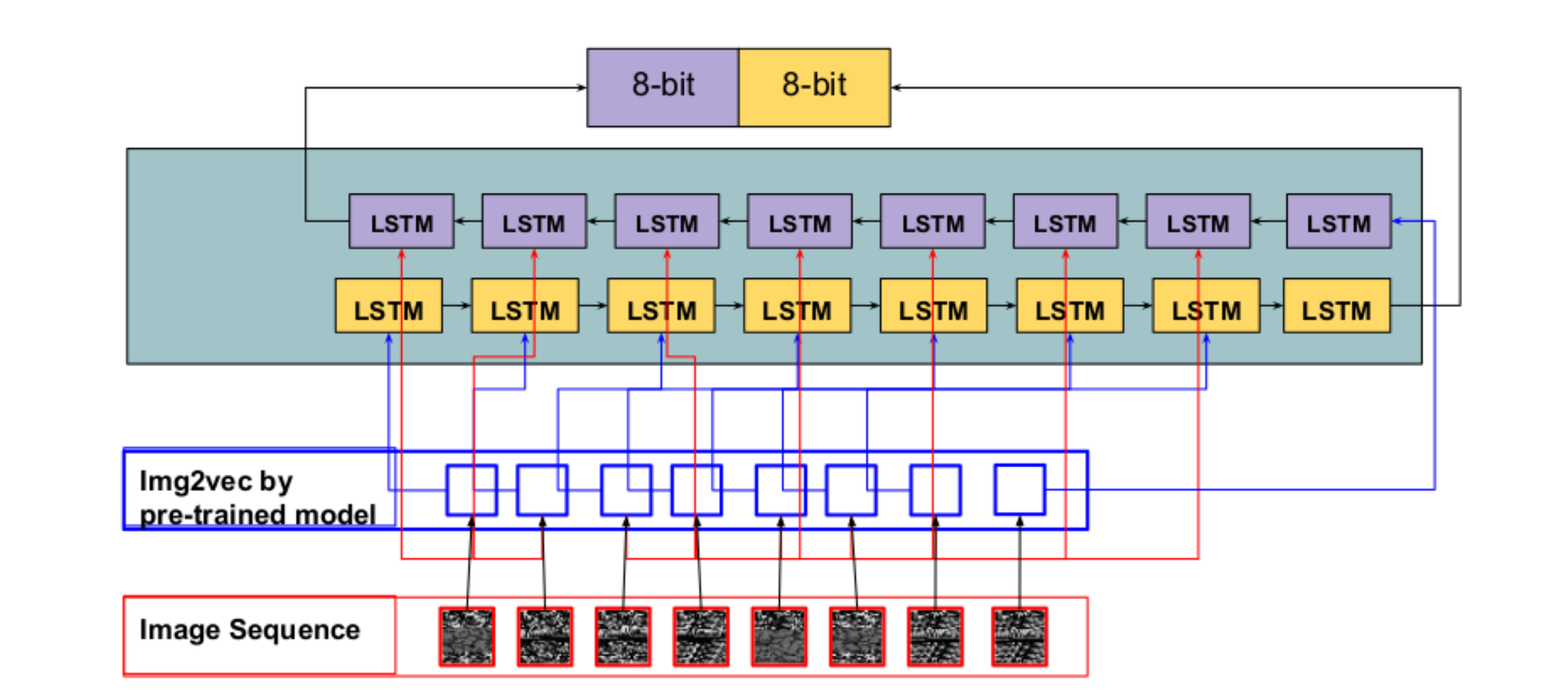}
	\caption{2D sequential image classification with bidirectional LSTM}
	\label{fig:lstm}
\end{figure}
\begin{figure*}
	\centering
	\includegraphics[width=0.6\textheight]{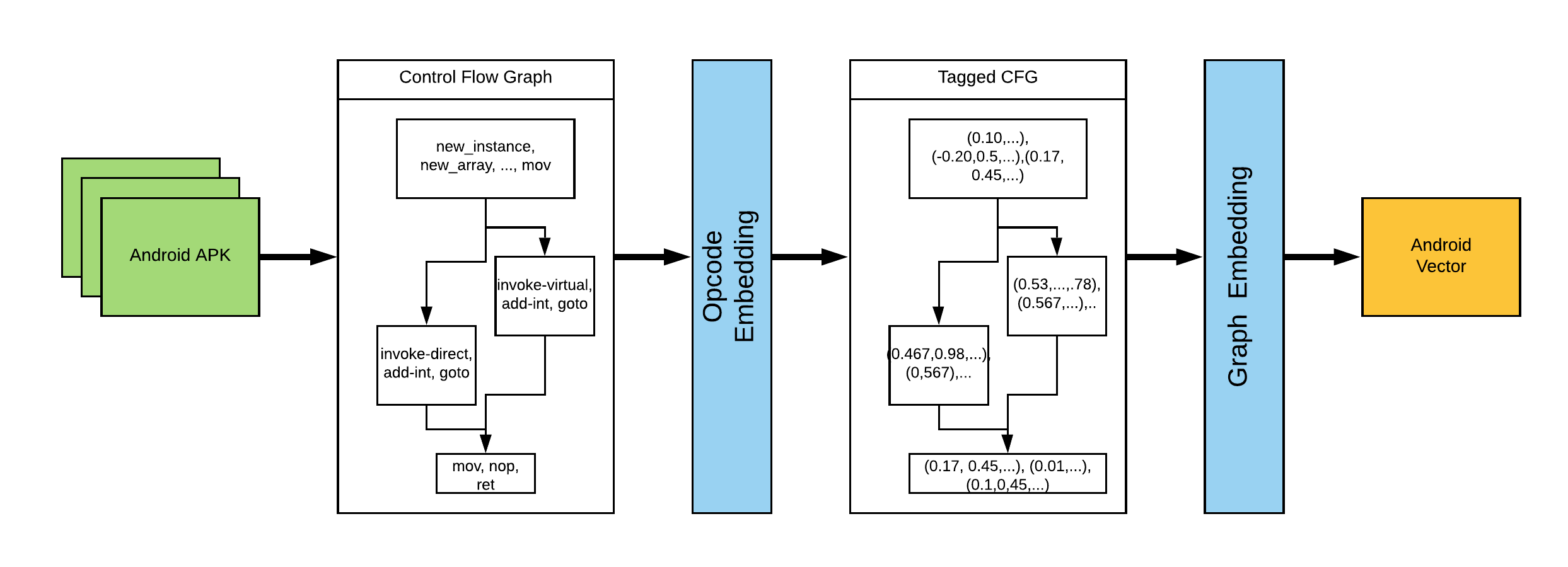}
	\caption{Converting Program Code to Vectors}
	\label{fig:graph2vec}
\end{figure*} 
Therefore, in our work to capture the network traffic's continuous characteristics, we introduce a bidirectional LSTM network on top of the pre-trained 2D-image classification model, which helps to extract image features from the converted network flows in the above section. This method can solve the \textbf{C2} effectively. Figure~\ref{fig:lstm} presents our sequential image classification structure. The above steps in aforementioned sections prepare the image sequences (Sec.~\ref{img_gen}) and img2vec (Sec.~\ref{img2vec}) model, which replaces each 2D gray image to 32-bit vector. Algorithm~\ref{algo:lstm} presents our processing flow. We use a bidirectional LSTM network, and the input of LSTM has converted vectors with $(1,32)$ shape. In Algorithm~\ref{algo:lstm}, both inputs for the forward and backward direction LSTM are the same. And $x_{t_{f}}$ stands for the input of the Forward LSTM and $h_{t-1}$ stands for the $t-1$ hidden state. $i_{t_f}$, $f_{t_f}$, and $o_{t_f}$ are input, forget and output gates, and $W^i_{f}$, $W^f_{f}$, $W^o_{f}$ are weight for the corresponding gates. For the Backward LSTM, those parameters have the same functionalities, but different values with them in Forward LSTM. At the initialize state, the input $x_{t_f}$ and $x_{t_b}$ are same. Furthermore, we concatenate the last hidden status $v_t$ as our final output vectors, where N stands for the number of all PCAP files. 
 
After getting the $v_t$ vectors for N PCAP files (N different Android samples), we use a full connection layer  followed a softmax layer to classify those raw network traffic to five different categories. We also use the sparse categorical cross-entropy to train our bi-LSTM network.

\subsection{Features from Code Graph Structure}
\label{code}
The overview of features from the code graph structure is illustrated with Figure~\ref{fig:graph2vec}.

\subsubsection{Opcode Embedding}
To simplify the procedure, we replace the instruction (opcode and operands) embedding with the opcode embedding. The reason for this replacement is as follows. First of all, the opcode represents Dalvik's instruction behaviors, and the operands represent the parameters. Dalvik's operands are virtual registers in a virtual machine. Those values are affected significantly by the undergoing usage of Dalvik VM or ART VM. Therefore, we cannot enumerate all of them. Additionally, if several malware samples in the same family use the same malicious pattern, the opcode itself can capture these malicious behaviors. 

In theory, our opcode embedding method may suffer from the \textit{operand removal} problem~\cite{haq2019survey}. One significant issue with that is that all the \textit{Invoke-Virtual} instructions\footnote{All the calling instructions such as \textit{invoke-super}, \textit{invoke-direct},\textit{ invoke-static}, and \textit{invoke-interface} suffer from the same problem.} have the same embedding vector, no matter what are the targets of the ~\textit{Invoke-Virtual} instruction.
For the opcode embedding, or \texttt{opcode2vec}, we map each opcode op$_i$ $\in$ OP (i.e., OP stands for the whole Dalvik opcodes) to a vector of the real number, using the \texttt{word2vec}~\cite{mikolov2013distributed} model with the skip-gram method. \texttt{word2vec} is an excellent feature learning technique, which is based on continuous bag-of-word and skip-gram techniques. The skip-gram technique aims to use the current opcode to predict the opcodes around it. We trained our \texttt{opcode2vec} model with a large corpus of opcodes extracted from the real applications.

\subsubsection{Function Embedding}

In this work, we treat the function embedding similar to the sentence embedding. 

We utilize the weighted mean of a non-empty finite multi-set of instruction's opcode to calculate the function embedding. Assuming the function $f$ includes n-opcode and a $l$-dimensional vector represents each opcode, the weight of the corresponding non-negative weights ${w_1,w_2,\dots,w_n}$ is:

\begin{equation}
\tilde{\vec{f}} = \frac{\sum_{i=1}^{n}{w_i x_i}}{\sum_{i=1}^{n}w_i}
\end{equation}
where $x_i$ represents the $l$-dimensional opcode embedding. 
%
%
%

The weighted mean function embedding is an easy and straightforward way. 

\subsubsection{Graph Embedding}
\begin{figure}[t]
	\centering
	\includegraphics[scale=0.15]{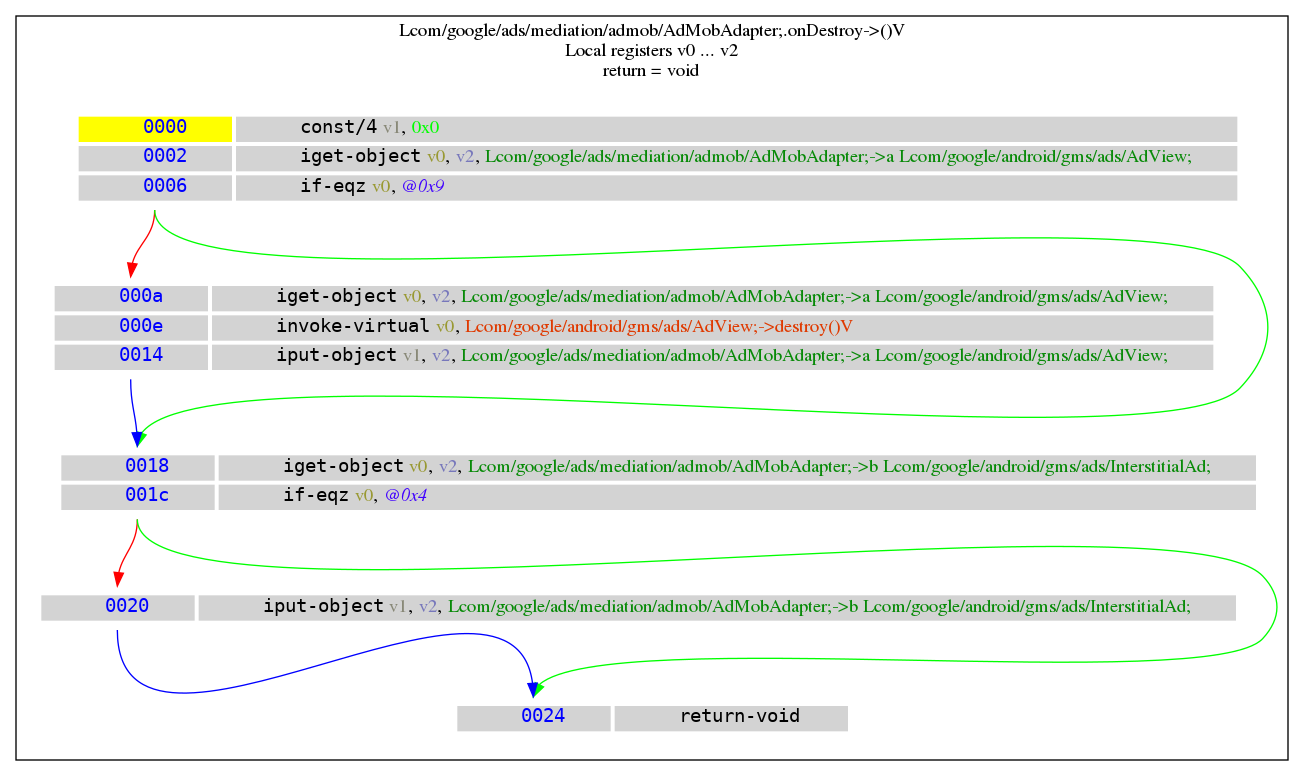}
	\caption{Function Call Graph}
	\label{fig:fcg}
\end{figure}

After getting the function embedding, we take those generated function embedding as the node embedding of the function call graph (FCG). In other words, we perform graph embedding on function call graph level. The function call graph is presented in Figure \ref{fig:fcg}, and it shows the FCG of an \textit{onDestroy()} function and several basic blocks insider of it. This module's ultimate purpose is to convert the graph representation to a vector and then take the generated vector as the neural network-based classifier's input.
We take structure2vec~\cite{dai2016discriminative} graph embedding method to convert one graph to an vector.

\begin{figure}

	\includegraphics[scale=0.20]{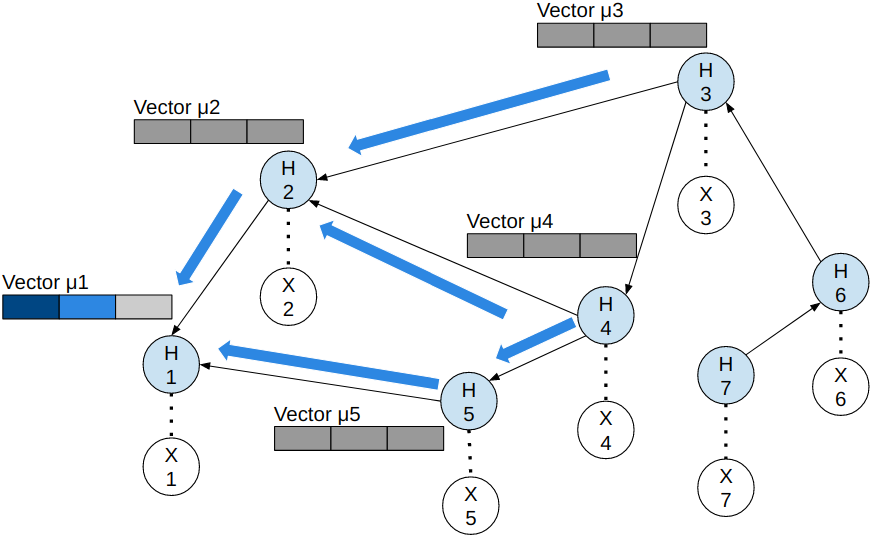}
	\caption{Graph Embedding}
	\label{fig:func2vec_graph}
\end{figure}

The technique at this step is presented in Figure~\ref{fig:func2vec_graph}. And we utilize the equations of (\ref{equ_6}), (\ref{equ_7}), (\ref{equ_8}) to convert a function call graph to a graph-vector, which stands for the whole Android application.
In our work, our graph-based function embedding includes two components. The first one is the control flow graph extraction, and the other one is the graph embedding for each function, which is adapted from the structure2vec.

In the real Android apps, each function/method is composed of multiple basic blocks. Figure~\ref{fig:func2vec_graph} exemplifies the graph embedding method to convert the whole graph to an vector.
Meanwhile, Figure~\ref{fig:func2vec_graph} presents the graph embedding, which is described in Algorithm~\ref{algo:ge}.

For the graph embedding, in our case, the vectors (nodes) of graphs are functions, and the edges are connections among those functions. Each vector (node) contains a set of opcodes inside it. The function embedding constructs each node's feature. Finally, a $p$-dimensional vector $\mu_i$ is associated with each vertex $v_i$.

We use adapted \texttt{structure2vec} to update the $p$-dimensional vector $\mu_i^{t+1}$ during the network training dynamically. The updating process is executed as follows:

\begin{equation}
\begin{aligned}
\mu_v^{(t+1)} = F(x_v, \sum_{u \in N_{v} }\mu_u^{(t)}), \forall v \in V. 
\end{aligned}
\label{equ_6}
\end{equation}

We initialize the $\mu_v^{(0)}$ at each vertex randomly. In practice, we design the function $F$ as the following form:

\begin{equation}
F(x_v, \sum_{u \in N_{v} }\mu_u^{(t)})=tanh(W_1x_v+\sigma(\sum_{u \in N(v)} \mu_u))
\label{equ_7}
\end{equation}

To make the nonlinear transformation $\sigma(.)$ effective, we define $\sigma(.)$ itself as an n layer fully-connected neural network.

\begin{equation}
\sigma(l) = P_1 * ReLU(P_2 * \dots ReLU(P_nl))
\label{equ_8}
\end{equation} 
The overall graph-based function embedding algorithm is illustrated with Algorithm~\ref{algo:ge}.

\begin{algorithm}
	\DontPrintSemicolon 
	\KwIn{Instruction embedding ${v_i : i\in I}$, control flow graph insider of a function $g_f$, parameter $\alpha$}
	\KwOut{Graph embedding ${v_f : f \in F}$}
	Initialize $\mu_v^{0} = \vec{Rand}, for all v \in V$
	
	\For{t=1 to T}{
		\For{$v \in V$}{
			$l_v = \sum_{u \in N(v)} \mu_u^{(t-1)}$
			
			$\mu_v^{(t)} = tanh(W_1x_v + \sigma(l_v))$
		}
	}
	$v_f$ = $W_2(\sum_{v \in V} \mu_v^{T})/len(V))$
	
	\Return{$v_f$}
	\caption{Graph embedding}
	\label{algo:ge}
\end{algorithm}

The graph embedding generates the vector embedding after all iterations. We use the average aggregation function as our last step to transform the vector embedding to the graph-based function embedding.

\subsection{Models training and Inference}

After getting our network traffic-based feature and also graph embedding for the function call graph,~\toolname concatenates two features together. And then we design a two-layer multi-layer layer perceptron (MLP) as our malware detection(binary classification) and malware categorization system(multi-classification, including Benign, Adware, Ransomware, Scareware and SMSmalware classes). In our network, malware detection is a binary classification issue. We label malware samples as ``1'' and benign samples as ``-1'' at the training step. During testing, we treat all the predictions less than zero as benign and the ones that are more than zero as malware.

\begin{equation}
\label{eq:detection}
f(G_h) = < (<g_i,w_{i1}> + b_{i1} ),w_{i2}> + b_{i2}
\end{equation}
where $w_{i1}, w_{i2}\in R^p$ is the weight of the 2-layer MLP network and $b_{i1}, b_{i2}\in R^p$ is the offset from the origin of the vector space. In this setting, a function call graph $G_h$ is classified as malicious if $f(G_h) > 0$
and as benign if $f(G_h) < 0$.

For the malware categorization, we divide this task into two sub-tasks. The first one categorizes those malware samples without pre-processing the malware samples. We label all the applications with a $N$-dimensional one-hot vector. The ``1'' in the one-hot vector stands for the index of the kinds of Android malware. We append one softmax layer, like Equation~\ref{eq:classification}, at the end of MLP and classify the malware to the classification ``n'', which stands for the type of malicious samples. In this task, we treat malware categorization as a multi-class classification issue.

\begin{equation}
\label{eq:classification}
f(G_h) = softmax(< (<g_i,w_{i1}> + b_{i1} ),w_{i2}> + b_{i2})
\end{equation}

On the other hand, we enumerate the top-n (n means the number of largest families) largest malware families as a pre-processing step and retrieve the malware dataset samples. If the sample is from the indicated malware family, we label it as ``1''. Otherwise, we label it as ``0''. Actually, with this assumption, we convert the multi-class classification problem to binary classification.

\section{Evaluation}
\subsection{Experimental Setup}
We set up our experiments on our server with 128GB RAM and 16 GB GPU. We train our model with Tensorflow2.0.0-beta0, Keras2.2.4, Sklearn0.20.0. We use SplitCap~\footnote{https://github.com/Master-13/SplitCap} tool to split PCAP files. Additionally, we use the pillow6.1.0 imaging library when we convert the network flows to images. Other assistant libraries are also used, such as numpy1.16.4 and matplotlib3.1.1. We use the Androguard~\footnote{https://github.com/androguard} to extract function call graphs from the Android APK files. 
\begin{table*}
	\centering
	\begin{tabular}{ccc}
		\hline
		Name & Number & Description \\
		\hline
		PCAP files & 2,126& All the raw network traffic files \\
		Network flows & 3,255,391& All network flows generated by methods in Section~\ref{network} \\
		Adware network flows & 580,170 & The Adware network flows partition \\
		Ransomware network flows & 382,279& The ransomware network flows partition     \\
		Scareware network flows & 517,954 & The Scareware network flows partition     \\
		SMSmalware network flows & 245,691& The SMSmalware network flows partition     \\
		Benign network flows &1,529,297& The network flows partition from benign applications \\
		\hline
		APK files &2,126& All program code files \\
		Adware apks &124&  No. of Android apks belong to Adware family \\
		Ransomware apks &112& No. of Android apks belong to Ransomware family \\
		Scareware apks &109& No. of Android apks belong to Scareware family\\ 
		SMSmalware apks &101& No. of Android apks belong to SMSmalware family\\
		Benign apks &1,700& No. of Android apks belong to benign \\
		\hline
	\end{tabular}
	
	\caption{The number of datasets, including network traffic and program code}
	\label{tab:dataset_number}
\end{table*}

\subsection{Dataset}
In our work, we collect two datasets. One is Android Malware CICMal2017~\cite{lashkari2018toward}, and the second dataset is AndroZoo~\cite{allix2016androzoo}.
For the Android Malware CICMal2017 dataset, it includes 426 malware and 1,700 benign samples and their corresponding network traffic raw files. IN our work, we consider CICMal2017 dataset to provide the dataset of dynamic feature. 

For the AndroZoo dataset, we collected samples following the suggestions of TESSERACT~\cite{pendlebury2019tesseract} and considered the spatial and temporal bias, and finally we set the collecting period between ~\textit{2020.03.31} and ~\textit{2021.03.31} and get in total 50,553 samples, which includes 25,553 malware and 25,000 benign samples. 

For the Malware detection issues, especially for the static feature-based work, the dataset with 2,126 samples is too small. However, for the dynamic feature-based work, 2,126 is the classic number. Actually, most works evaluate those frameworks with smaller datasets, such as~\cite{jeon2020dynamic} evaluates on 1,401 samples (1,000 Malware and 401 Benign). \cite{onwuzurike2018family} takes 2,336 benign and 1,892 malware samples, and the total number is 4,228. Because our network flow-based dataset is captured when the Android application runs, that is a classic dynamic feature. The number of the dataset for dynamic analysis (\cite{jeon2020dynamic,onwuzurike2018family}) is normal, which is smaller than static analysis (~\cite{ye2019out} evaluates on a dataset with 83,784 benign and 106,912 samples based on static analysis, ~\cite{xu2020manis} evaluates on the 49,947 benign and 5,560 malware based on static analysis).

In addition, the number of various categories is illustrated in Table~\ref{tab:dataset_number} in detail. For the network traffic, we extract 3,255,391 network flows in total from 2,216 PCAP files. Here, to pre-train our 2D gray image classification task, we create our dataset,~\tooldataset, which provides a benchmark to network traffic analysis with the convolution neural network. We split the dataset with 80\% training and 20\% testing in our experiment.
\subsection{Power Law and Opcode Embedding}
	\begin{figure*}
	    \centering
        \includegraphics[scale=0.85]{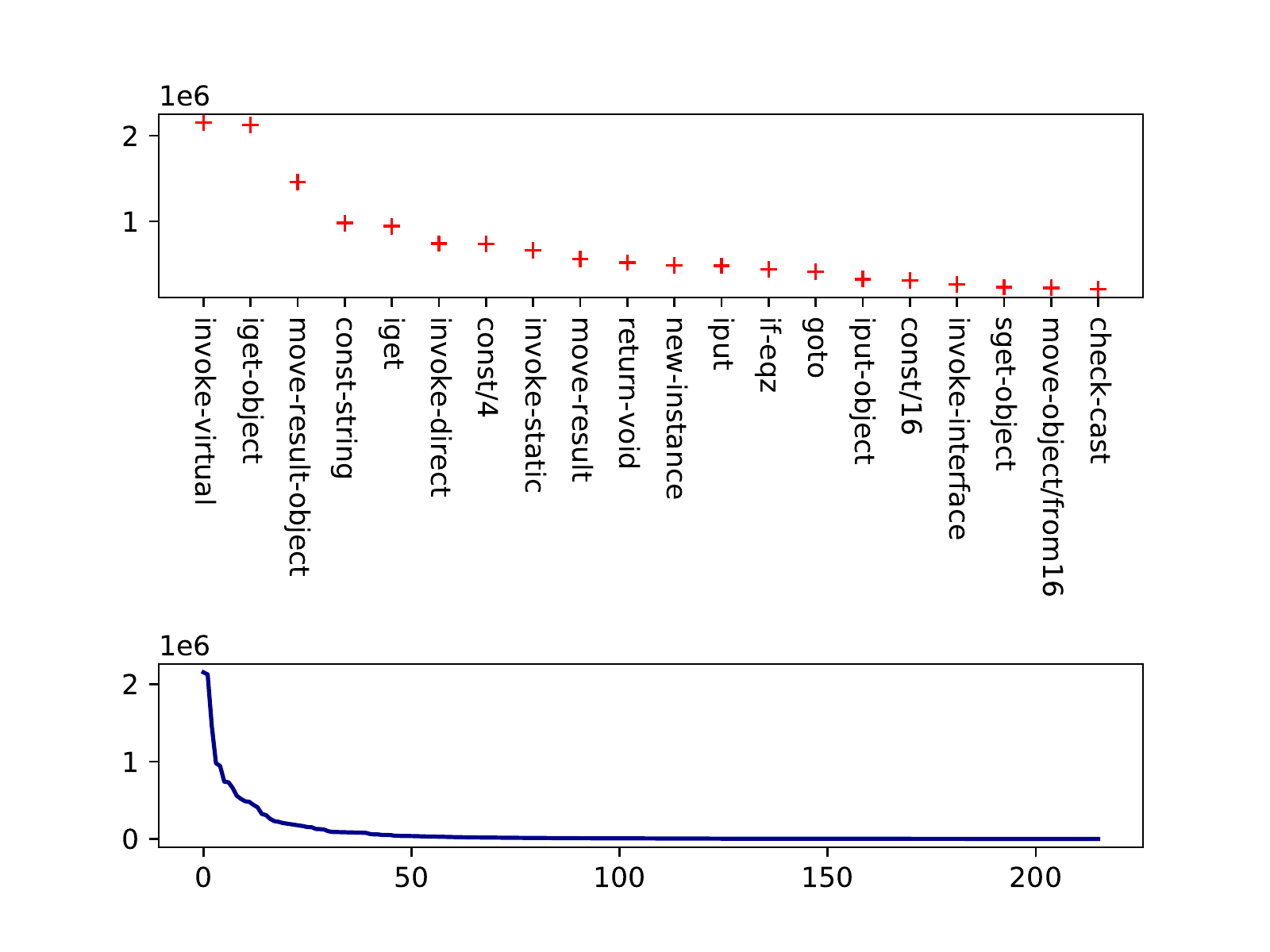}
		\caption{Power-law Distribution for Dalivk Opcodes.}
		\label{fig::powerlaw}
	\end{figure*}

Before moving to our evaluation tasks, we use the distribution of our opcode to prove the reason-ability of using natural processing language techniques in our works. In order to get the reasonable opcode2vec module, we pre-train the opcode2vec by the AndroZoo dataset. We extract all opcode by Androguard tool and obtain 18,240,542 opcodes in total. And then we take those opcodes as our word corpus to train the opcode2vec model.  Figure~\ref{fig::powerlaw} presents the opcode distribution for the above datasets. Figure~\ref{fig::powerlaw} shows Dalvik's opcode distribution, which has 216 opcodes and top-20 opcodes are presented.  All of them follow the power-law distribution, which means borrowing word embedding techniques from natural language processing to do opcode embedding is reasonable. 
\subsection{Results Comparison}

\begin{table*}
	\centering
	\begin{tabular}{ccccc}
		\hline
		Classifier&Accuracy(\%)&Precision(\%)&Recall(\%)&F1(\%)  \\ \hline
		Drebin~\cite{arp2014drebin} &96.58&95.37&97.85&96.59 \\
		Adagio~\cite{gascon2013structural} &95.01&91.07&100&95.32 \\
		Droidmat~\cite{wu2012droidmat} &89.87&90.89&88.28&89.56 \\
		CICAndMal2017~\cite{lashkari2018toward} &95.77&96.00&95.64&95.77 \\ \hline
		\toolname-NET(bi-LSTM) &95.39&95.54&95.39&95.18 \\ 
		\toolname-Graph &95.18&92.31&83.72&87.80 \\ \hline
		\toolname &97.16&97.13&97.16&97.09 \\ \hline
		
	\end{tabular}
	\caption{Results Comparison for the Malware Detection Task}
	\label{tab:binaryclassification}
\end{table*}
\begin{table*}
	\centering
	\begin{tabular}{ccccc}
		\hline
		Classifier&Accuracy(\%)&Precision(\%)&Recall(\%)&F1(\%) \\
		\hline
		CICAndMal2017~\cite{lashkari2018toward} &86.85&85.92&86.85&84.82 \\ \hline
		\toolname-NET(bi-LSTM) &84.70&80.22&84.70&82.39 \\ 
		\toolname-Graph &88.45&100.0&93.87&88.45 \\ \hline
		\toolname &88.32&85.82&88.32&87.05 \\ \hline
	\end{tabular}
	\caption{Results Comparison for the Malware Categorization Task(average = 'weighted')}
	\label{tab:multiclassification}
\end{table*}

This section compares our results with other related works, both from the program code and network traffic-based works.
Besides presenting the hybrid pattern-based malware detection and categorization, we also present our generated results by only considering the network traffic and program code features. Additionally, we also re-implement ( Droidmat~\cite{wu2012droidmat}) and reproduce (Drebin~\cite{arp2014drebin}~\footnote{https://github.com/alisakhatipova/Drebin}, Adagio~\cite{gascon2013structural}~\footnote{https://github.com/hgascon/adagio}) other related works and compare them with our framework (the results of those frameworks have a little difference with the original works because of the different dataset.). For our own ~\toolname, after preparing the dataset as CSV files, we use the RandomForest(RF) classifier by default to do our malware detection and malware categorization. Our RF is defined as 1,400 trees in the forest and 80 as the tree's maximum depth. We set $min\_samples\_split$ as 5 and the number of features to consider when looking for the best split as \text{sqrt}.

The motivation of our work is to detect malware by combing the static (program code) and dynamic (network traffic) features. If we did not combine those features, how
about the malware detection performance if we only consider the program code or network
traffic? We evaluate this issue in Table~\ref{tab:binaryclassification}. HP-MDC-NET(Wang et al.) and HP-MDC-NET(bi\_LSTM) are introduced to evaluate the performance of network traffic, and HP-MDC-Graph is presented to evaluate the performance of program code.

\begin{table*}
	\centering
	\begin{tabular}{cc}
		\hline
		Classifier & Description \\ 
		\hline
		RF & n\_estimators=1400, min\_sample\_split=5,max\_features="sqrt",max\_depth=80 \\
		AdaBoost & All default values \\
		GradientBoost & lr=0.01, n\_estimators=1500, max\_depth=4, min\_samples\_split=40 \\ 
		MLP & sover='sgd',alpha=1e-5,hidden\_layers\_sizes=(400,400,200,100,10) \\
		DecisionTree & min\_samples\_split=10,max\_features='sqrt',max\_depth=20 \\
		\hline
		
	\end{tabular}
	\caption{Various classifiers setting}
	\label{sec:setting}
\end{table*}

\begin{table}
	\centering
	\begin{tabular}{ccccc}
		\hline
		Classifier & Accuracy & Precision & Recall & F1\\ \hline
		RF &97.16&97.13&97.16&97.09 \\
		AdaBoost &93.13&92.81&93.13&92.85 \\
		GradientBoost &96.88&96.83&96.88&96.80 \\
		MLP &91.01&90.48&91.01&90.02 \\
		DecisionTree &93.66&93.64&93.66&93.65 \\ \hline
	\end{tabular}
	\caption{\toolname's Performance with various classifier}
	\label{various}
\end{table}
For the image-based embeddings, actually, we provide two compared results. The first one is HP-MDC-NET(Wang et al.) in Table~\ref{tab:binaryclassification}. And the other one is HP-MDC-NET(bi\_LSTM) in
Table~\ref{tab:binaryclassification}. Both experiments are using image-based embeddings and convolutional neural network to do malware detection. The latter appends a bi-directional LSTM network to obtain the final vectors. Meanwhile, the former is the re-implementation of \cite{wang2017malware}. We create our own dataset -~\tooldataset and evaluate against HP-MDC-NET(Wang et al.) and HP-MDC-NET(bi\_LSTM). In contrast with our ~\toolname, Wang et al.\cite{wang2017malware} can only classify each network-flow to malware or benign. However, ~\toolname can do the network-flow and network-packet level classification. To classify each network-flow as the malware or benign is not our purpose because we need network-packet classification to clarify an application is malware or benign, and thus we design and evaluate the other method HP-MDC-NET(bi\_LSTM).

Table~\ref{tab:binaryclassification} illustrates the malware detection (binary classification) performance.
Table~\ref{tab:binaryclassification} shows that \toolname-Net(Wang et al.) gets the best performance, which catches up to 98\% accuracy. However, this experiment processes the malware classification on~\tooldataset similar to the digital handwriting classification on the MNIST dataset, which indirectly did the classification. That means we firstly extract and convert all network flows to images and then classify all images belong to one class. For example, for a PCAP file, e.g., \textit{ad-ewind-koodous-1de012.pcap}, we extracted and converted network flows to images and got 3,329 samples. The 98\% accuracy means 98\% of 3,329 samples are classified correctly. However, we cannot determine the whole network flows characteristics because mostly malicious behaviors are hidden in a few network flows by sophisticated attackers. Even if we get high performance of over 98\%, we cannot infer that this classifier is the best one. Therefore, we introduced \toolname, which combines network traffic and program code patterns. With this method, ~\toolname gets 97.16\% accuracy(except  \toolname-Net(Wang et al.), we analyze the \toolname-Net(Wang et al.) above). In contrast to other methods, our~\toolname gets better results than others. The results are illustrated in Table\ref{tab:binaryclassification} in detail.

In our experiment, the malware categorization task is a multi-class classification issue. Similar to the malware detection (binary classification) task, \toolname-NET on~\tooldataset gets the best performance on the image classification task indirectly. Take the same example with malware detection above, 97.23 accuracy means 97.23\% images from \textit{ad-ewind-koodous-1de012.pcap} are classified to Adware class. However, we cannot determinedly infer this PCAP is Adware network traffic. Additionally, we compare our results only with CICAndMal2017 because most Android malware detection works, such as Drebin, Adagio, and Droidmat did not consider the malware categorization. Although FM~\cite{li2018android} considers the malware categorization task, it converts the multi-class task to binary-class (If one malware sample belongs to a specific malware family, then the label is 1; otherwise, that is 0.).  Our ~\toolname on the multi-class classification task gets better results than CICAndMal2017. The primary reason is that essential patterns for various malware families represent the manually indicated features by CICAndMal2017 losing information. Our method can catch up better malware families' features by representation learning. CICAndMal2017 uses the
additional tools to extract nine features from network PCAP file and detect malware with those
extract features. And our results show that image-based method is better than rule-based
methods. 
Table~\ref{tab:multiclassification} presents the performance results of the malware categorization task.

Last but not least, besides the RandomForest (RF) classifier, we also consider the other four classifiers for malware detection and categorization. The classifiers are described in Table~\ref{sec:setting}. Besides the setting in Table~\ref{sec:setting}, we use all default parameters. Due to the limited space, we only present results in Table~\ref{various} for the malware detection (binary classification) task. Table~\ref{various} shows that the RF classifier gets the best performance and then followed by the GradientBoost classifier. MLP gets the worst in our ~\toolname framework.

\section{Related Works}
\subsection{Static Feature based Android Malware Detection}
Recently, the area of Android malware detection has attracted several security researchers . In this section, we first discuss related work on the static analysis of Android malware, then proceed to present approaches related to Android Malware detection using structural analysis.

Li et al.~\cite{li2018android} present a factorization machine(FM) based Android malware detection system. It extracts seven types of features that are
highly relevant to malware detection from the manifest file and
source code of each mobile application, including Application Programming Interface (API) calls, permissions, etc. They propose to use the factorization machine, which fit the problem the best, as a supervised classifier for malware detection. According to extensive performance evaluation, our proposed method achieved a test result of 99.01\% detection rate with a false positive rate of 0.09\% on the DREBIN dataset. To contrast with it, we do not use permissions and API calls as our features although both the FM and our solution get over 99\% performance. We extract function call graph from Android apps, and construct the graph neural network based detection system. Our solution is not sensitive to the obfuscation techniques, especially the string-based obfuscation (e.g., function names, variable names). In addition, both FM and our work consider the malware detection and malware classification (top-n families) problems. One difference is that we also consider the malware classification on the whole malware families perspective, which FM doesn’t consider.

Adagio~\cite{gascon2013structural} shows a kernel-hashing based malware detection system on the function call graph. It is based on the efficient embeddings of function call graphs with an explicit feature map inspired by a linear-time graph kernel. In an evaluation with real malware samples purely based on structural features, Adagio outperforms several related approaches and detects 89\% of the malware with few false alarms, while also allowing to pin-point malicious code structures within Android applications. Both the Adagio and our solution are based on the function call graph of Android applications. However, we design the graph embedding based on the function call graph, while Adagio uses the kernel-hashing method. In addition, we use opcode embedding to transfer the opcodes of Dalvik's instructions to vectors while Adagio uses a reduced one-hot vector to represent the opcode. There is also an over-approximation problem in the Adagio because they narrow down the 221-opcode to 15-opcode.

There are also many works based on the n-gram opcode to detection Android malware, such as~\cite{mclaughlin2017deep,canfora2015effectiveness}.
McLaughlin et al.~\cite{mclaughlin2017deep} propose a novel Android malware detection system with the convolutional neural network(CNN). The network automatically learns features of malware from the raw opcode sequences and thus removing the need for hand-engineered malware features. Meanwhile, Canfora et al. ~\cite{canfora2015effectiveness} investigate that if the frequencies of n-grams of opcodes are useful in detecting Android malware and if there are some significant malware families for which they are more or less effective. To this end, They design a method based on state-of-the-art classifiers applied to frequencies of opcodes. Results show that an accuracy of 97\% can be obtained on average. To contrast with our work, we only use the frequencies of n-grams of opcodes in our SIF-function embedding method. In addition to that, we also have the graph neural network to convert the whole Android application to a vector, which considers the 'Function Jumping' issues while the above approaches handle the Android application as a sequence of opcodes without considering 'Function Jumping'. 

Program code-based malware detection methods extract features from the code itself. Technically those features include the API calls, N-gram, and control flow graph (CFG) based methods.  API call based malware detection~\cite{aafer2013droidapiminer,wu2012droidmat,peiravian2013machine,li2018android,arp2014drebin} uses API calls to detect Android malware.  
N-gram-based Malware Detection is based on the n-gram opcode to detect Android malware, such as~\cite{mclaughlin2017deep,kang2016n,raff2017malware}. Last but not least, Graph-based malware detection systems use graph structure to do detection tasks, such as Apk2vec~\cite{narayanan2018apk2vec} and Adagio~\cite{gascon2013structural}.

\subsection{Dynamic feature based Android Malware Detection}
Machine learning and deep learning techniques are heavily introduced into the network traffic analysis. Researchers use manual indicated features to recognize network traffic patterns with traditional machine learning algorithms, such as traffic classification, network security, and anomaly detection~\cite{wang2017malware,marin2020deepmal,lashkari2018toward,taheri2019extensible,arora2014malware}.  Those features include network ports, deep packets inspection, statistical, and behavior-based features. To contrast with the above works, our~\toolname considers the network traffic as 2D-images sequences and using CNN and LSTM based neural networks to represent the network traffic information. That could be more accurate than port and other rule-based methods.

\section{Limitation}
To contrast with other machine learning and deep learning-based works in the malware detection field, on the one hand, our \toolname can catch up not only syntax (such as function call graph connection) information but also the semantic information (such as the network traffic with flow representation). However, the whole process is more time-consumption than rule-based methods, such as port-based malware detection with network traffic and permission-based Android malicious program detection. 

On the other hand, in contrast to other Android malware detection works, the dataset, especially the program code-based dataset, is too small. Although our evaluation demonstrates better performance than its precedent, we need to increase the number of samples in the future.

Last but not least, in our work we did not consider the adversarial machine learning because that is not a simple issue, especially for
the graph-based features because of the non-continuous feature. MANIS~\cite{xu2020manis} only
discusses how to evade the malware detection system on graphs. In that work, one adversarial machine learning based method to evade the malware detection system on graphs is described in detail. The graph structure in that
paper is significantly easier than the graph structure and natural language processing inspired
methods in this paper. Therefore, to consider the adversarial learning for our hybrid-feature
malware detection system, we plan continue this as our future work.
\section{Conclusion}
In this work, we present~\toolname, a hybrid pattern malware detection and categorization framework. We use the transfer learning method to extract features from the network traffic and program code with the pre-trained models. We treat it as a 2D gray image sequence classification task for the network traffic-based classification and use a bi-directional LSTM to process image sequences. For the 2D gray image, we use an 8-layer CNN to pre-train the gray images, which stand for the network flows. For the program code part, we use the \texttt{opcode2vec} and \texttt{function2vec} to convert the functions in machine instruction to vectors and then use graph embedding to obtain the final vectors, which stand for the Android applications. We boost our malware detection and categorization performance with the hybrid pattern, and our results outperform most related works. 

\small{
\bibliographystyle{unsrt}
\bibliography{paper}
}
\end{document}